%% file: ms.tex
\documentclass[runningheads]{llncs}
\usepackage{graphicx}
\usepackage[english]{babel}
\usepackage[utf8]{inputenc}
\usepackage{booktabs}
\usepackage{makecell} 
\usepackage{arydshln}
\usepackage{amsmath}
\usepackage[caption=false,font=footnotesize]{subfig}
\usepackage[export]{adjustbox}
\usepackage{xcolor}
\usepackage[misc]{ifsym}
\def\Plus{\texttt{+}}

\usepackage{color}

\usepackage{graphicx}
\usepackage{hyperref}

\usepackage[inline]{enumitem}

\newcommand{\SKIP}[1]{}

\begin{document}
\title{Double Encoder-Decoder Networks \\ for Gastrointestinal Polyp Segmentation}
\titlerunning{Double Auto-Enconders for Polyp Segmentation}

\author{Adrian Galdran\inst{1} \and
Gustavo Carneiro\inst{2} \and
Miguel A. Gonz\'alez Ballester\inst{3,4}}
\authorrunning{A. Galdran et al.}
\institute{Dpt. of Computing and Informatics, Bournemouth University, UK, \email{agaldran@bournemouth.ac.uk} \and
Australian Institute for Machine Learning, University of Adelaide, Australia \and
BCN Medtech, Dept. of Information and Communication Technologies, \\Universitat Pompeu Fabra, Barcelona, Spain \and
ICREA, Barcelona, Spain}

\maketitle              %
\begin{abstract}
Polyps represent an early sign of the development of Colorectal Cancer. 
The standard procedure for their detection consists of colonoscopic examination of the gastrointestinal tract. 
However, the wide range of polyp shapes and visual appearances, as well as the reduced quality of this image modality, turn their automatic identification and segmentation with computational tools into a challenging computer vision task.
In this work, we present a new strategy for the delineation of gastrointestinal polyps from endoscopic images based on a direct extension of common encoder-decoder networks for semantic segmentation. 
In our approach, two pretrained encoder-decoder networks are sequentially stacked: the second network takes as input the concatenation of the original frame and the initial prediction generated by the first network, which acts as an attention mechanism enabling the second network to focus on interesting areas within the image, thereby improving the quality of its predictions. 
Quantitative evaluation carried out on several polyp segmentation databases shows that double encoder-decoder networks clearly outperform their single encoder-decoder counterparts in all cases. 
In addition, our best double encoder-decoder combination attains excellent segmentation accuracy and reaches state-of-the-art performance results in all the considered datasets, with a remarkable boost of accuracy on images extracted from datasets not used for training.
\end{abstract}

\input{1_introduction}

\input{2_method}

\input{3_results}

\input{4_discussion}

\bibliographystyle{splncs04}
\bibliography{aiha_polyp.bib}

\end{document}

%% file: 1_introduction.tex
\begin{figure*}[t]
\centering
\subfloat[]{\includegraphics[width = 0.24\textwidth,valign=c]{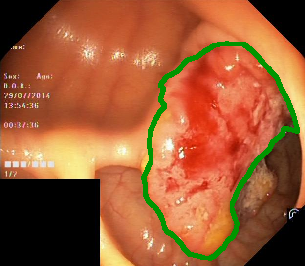}
\label{fig_data_1}}
\hfil
\subfloat[]{\includegraphics[width = 0.24\textwidth,valign=c]{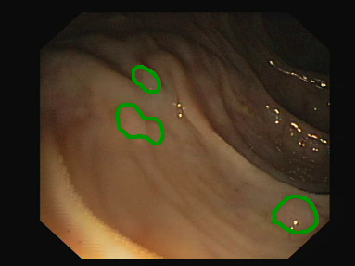}
\label{fig_data_2}}
\hfil
\subfloat[]{\includegraphics[width = 0.24\textwidth,valign=c]{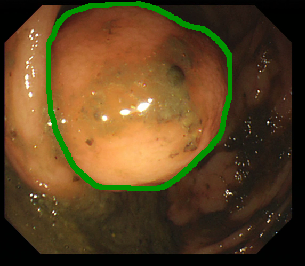}
\label{fig_data_3}}
\hfil
\subfloat[]{\includegraphics[width = 0.24\textwidth,valign=c]{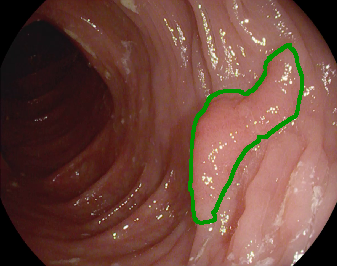}
\label{fig_data_4}}
\caption{Polyp visual aspects have a wide variety in terms of shape and color. Four different polyps sampled from the different databases considered in this work: (a) Kvasir-Seg \cite{jha_kvasir-seg_2020}, (b) CVC-ClinicDB \cite{bernal_wm-dova_2015}, (c) CVC-ColonDB \cite{bernal_towards_2012}, (d) ETIS \cite{silva_toward_2014}.}
\label{fig_datasets}
\end{figure*}

\section{Introduction}
The large bowel within the human gastrointestinal tract can be affected by different diseases, among which, Colorectal Cancer (CRC) is particularly concerning. 
CRC represents the second most common cancer type in women and third most common for men~\cite{haggar_colorectal_2009}. 
Gastro-intestinal polyps are known precursors of this type of cancer~\cite{gao_benchmark_2017}, being present in almost half of the patients over 50 undergoing screening colonoscopies~\cite{sanchez-peralta_deep_2020}. 
This kind of lesions show a wide range of shapes and visual appearances, as shown in Fig.~\ref{fig_datasets}, turning its identification and segmentation into a challenging problem.

Treatment of CRC starts with the detection of colorectal paraneoplastic lesions, performed during colonoscopy screenings. 
In this endoscopic procedure, a flexible tube with a light camera mounted on its tip is introduced through the rectum to find and sample (or resect) polyps from the colon. 
Early detection of CRC has been demonstrated to substantially increase 5-year survival rates, with screening programs enabling even pre-symptomatic treatment \cite{sanchez-peralta_deep_2020}. 
Unfortunately, it is estimated that around 6-27\% of polyps are missed during a colonoscopic examination~\cite{ahn_miss_2012}.
It has been recently shown in~\cite{lui_new_2020} that up to 80\% of missed lesions could be avoided with real-time computer-aided colonoscopic image analysis and decision support systems. 
Therefore, computer-aided polyp detection has been extensively explored as a complementary tool for colonoscopic procedures to improve detection rates, enable early treatment, and increase survival rates. 

The most relevant computer-aided tasks related to polyp analysis in endoscopic imaging are:
\begin{enumerate*}
\item Polyp Detection: Deciding if polyps appear in an endoscopic frame~\cite{bernal_comparative_2017}.
\item Polyp Classification: Assigning polyps to a range of sub-categories or degrees of malignancy~\cite{carneiro_deep_2020}.
\item Polyp Localization: finding the position (usually in terms of a bounding box) of polyps within a frame~\cite{zhang_polyp_2018}. 
\item Polyp Segmentation: delineating the exact polyp contour in a given endoscopic frame~\cite{wickstrom_uncertainty_2020}.  
\end{enumerate*}

This work is concerned with the task of polyp segmentation, which has attracted much attention in recent years. 
Polyp segmentation is typically achieved by means of encoder-decoder architectures composed a pair of Convolutional Neural Networks (CNN). 
The main variation across different works is the design of such architectures. For instance in \cite{guo_polyp_2020} an encoder-decoder network containing multi-resolution, multi-classification,
 and fusion sub-networks was proposed, and in \cite{sanchez-peralta_eigenloss_2020} several combinations of different encoder and decoder architectures were explored. 
In \cite{fang_selective_2019} an architecture in which there is a shared encoder and two mutually depending decoders that model polyp areas and boundaries respectively is introduced,  whereas in \cite{kang_ensemble_2019} ensembles of instance-segmentation architectures like Mask-RCNN were studied, and in \cite{fan_pranet_2020} parallel reverse attention layers are proposed to better capture the relationship between polyp areas and their boundaries.
Two-stage detection/segmentation pipelines, in which a first an object detector roughly locates a polyp and then an encoder-decoder CNN delineates its margins have also been explored, \textit{e.g.} in \cite{jia_automatic_2020}.
Alternative loss function have also been studied, \textit{e.g.} in \cite{sanchez-peralta_eigenloss_2020}.
A recent review of these and other approaches can be found in \cite{sanchez-peralta_deep_2020}.

\begin{figure*}[t]
\centering
\includegraphics[width = \textwidth]{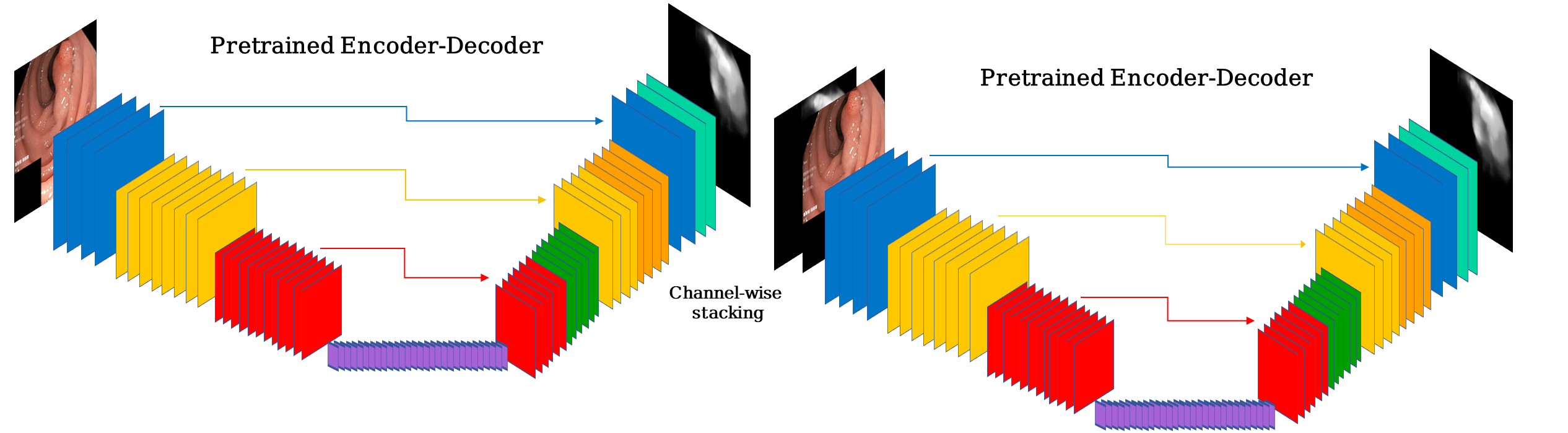}
\label{fig_deg_1}
\caption{Pre-trained Double Encoder-Decoder Network for Polyp Segmentation. The second network receives as input the original colonoscopic frame concatenated with the prediction of the first network; we conjecture that this allows the second network to better focus on interesting areas of the image, improving its segmentation accuracy.}
\label{fig_enc_dec}
\end{figure*}

The approach proposed here consists of stacking two segmentation networks in a sequential manner, where the second network receives as input the concatenation of the prediction from the first one with the original frame, as shown in Fig.~\ref{fig_enc_dec}. 
This way, the output of the first network acts as an attention mechanism that provides the second network with a map of interesting locations on which the second network should focus. 
It should be noted that iterative segmentation networks have been proposed in other contexts recently, such as the method by Li et al.~\cite{li_iternet_2020}, which proposes stacking a large U-Net with $N-1$ smaller U-Nets for the task of retinal vessel segmentation. 
The main difference compared with our technique is that in~\cite{li_iternet_2020} the subsequent small U-Nets do not process the input image, but rather the candidate segmentation only. 
Also for retinal vessel segmentation, the recent paper by Galdran et al.~\cite{galdran_little_2020} makes use of two consecutive U-Nets, but their focus is on building compact models for a substantially simpler task, not benefiting from using pre-trained CNN. 
The Double-UNet, introduced in~\cite{jha_doubleu-net_2020}, holds also some similarities with our proposed approach, but in their approach the output of the first network is multiplied by the input image (instead of concatenated), potentially discarding useful information, which may not be recovered, for the second network. In addition, the backbone encoder for Double-Unet is a VGG-19~\cite{simonyan_very_2015}, which is currently not competitive with the state-of-the-art classification networks that have been proposed in recent years.

In this work, we propose a framework for semantic segmentation based on the sequential use of two encoder-decoder networks. 
Extensive experimental results with three different pre-trained encoders and decoders on a recently proposed polyp segmentation dataset \cite{jha_kvasir-seg_2020} show that our double encoder-decoder networks provide a boost in performance over their single encoder-decoder in all cases. 
In addition, results across different combinations of encoders show substantial robustness and consistency.
Our best model is re-trained on a common polyp segmentation benchmark in order to enable comparison with the state-of-the-art, attaining results that considerably improve upon several recent approaches, with remarkable performance on three unseen datasets, where our approach achieves a performance boost in terms of Dice score of 1.52\%, 5,29\%, and 12,07\% when compared to the second best considered method.

%% file: 2_method.tex
\section{Methodology}

\subsection{Double Autoencoders}
Dense semantic segmentation tasks that produce per-pixel predictions are typically approached with an encoder-decoder network \cite{badrinarayanan_segnet_2017,tian_decoders_2019,wojna_devil_2019}. 
In this context, the encoder can be regarded as a feature extractor that downsamples spatial resolution and increases the number of channels by learning convolutional filters. 
After the image has been encoded into a low-dimensional representation, the decoder upsamples this representation back to the original input size, thereby generating a pixel-wise prediction. 
Typically, skip connections are added to map information from the encoder to the decoder \cite{ronneberger_u-net_2015}. 
Double encoder-encoders are a direct extension of encoder-decoder architectures in which two encoder-decoder networks are sequentially combined.
Denoting by $x$ the input RGB image, $E^{(1)}$ the first network, and $E^{(2)}$ the second network, in a double encoder-decoder, the output $E^{(1)}(x)$ of the first network is provided to the second network together with $x$ so that it can act as an attention map that allows $E^{(2)}$ to focus on the most interesting areas of the image:
\begin{equation}\label{wnet_def}
E(x) = E^{(2)}(x, E^{(1)}(x)),
\end{equation}
where $x$ and $E^{(1)}(x)$ are stacked so that the input to $E^{(2)}$ has four channels instead of the three channels corresponding to the RGB components of $x$. 
This is illustrated in Fig. \ref{fig_enc_dec}.
There are some choices to be made in this framework, specifically about the structure of the encoder and decoder sub-networks within $E^{(1)}$ and $E^{(2)}$. 
Note that $E^{(1)}$ and $E^{(2)}$ do not need to share the same architecture, although in this work we restrict ourselves to this case to simplify the exposition.
In the next two sections we describe different alternatives that we will thoroughly explore in our experiments.

\subsection{Pretrained Encoders}\label{encoders}

Double encoder-decoder networks contain two sub-networks $E^{(1)}$ and $E^{(2)}$, each of which have an encoder branch that can be chosen as any Convolutional Neural Network designed for classification purposes. 
Image representations obtained at the end of this branch are expected to be useful to classify objects within the image and to segment objects when iteratively upscaled in the decoder branch. 
In particular, it is favorable to select top-performer architectures that have been shown to work well for natural image classification tasks, since we can then re-use the weights of the decoder. 
This experimentally leads to faster convergence, requires less training data and leads to better generalization. 
In this work we consider three such architectures, which we briefly review below. 
These range from more compact models to larger, more powerful architectures, to better illustrate the benefits of double encoder-decoder networks under different scenarios.

\subsubsection{MobileNet V2}
MobileNet is a mobile architecture specifically designed to maximize compactness, speed, and efficiency~\cite{sandler_mobilenetv2_2018,sandler_mobilenetv2_2018} to enable its deployment on mobile devices. 
The main feature of this network is that it implements efficient depthwise separable convolutions to reduce computational load.
Later improved versions of this architecture adopt ``inverted residual'' blocks on which shortcut connections are established between ``thin'' activation volumes that contain less features, considerably reducing memory constraints. 
In addition, interleaved non-linearities within narrow layers of the network are avoided, which is shown to further decrease computational demands.

\subsubsection{Resnet34}

Residual Network architectures (ResNets) were first introduced in \cite{he_deep_2016}, and have since then become the default networks for Computer Vision tasks, due to their excellent complexity versus performance trade-off.
ResNets popularized the notion of skip-connections, which were the main ingredient of residual blocks: a series of convolutions are applied to the input of a particular network layer, but a by-pass is added to the input so that information can flow unaltered up to the output of the block. 
This improves the training process of very deep networks, helping to avoid vanishing gradient problems. 
Although large ResNet architectures with up to 152 layers have been proposed, as well as improved versions like ResNext \cite{xie_aggregated_2017}, we limit our experiments to a relatively small ResNet variant, namely ResNet34. 
This provides a nice intermediate stage in between the compactness of MobileNet and the complexity of Dual-Path-Networks, which we describe next.

\subsubsection{Dual-Path Networks}

Dual Path Networks (DPNs) were presented in \cite{chen_dual_2017}, and they provide an hybrid design that merges different aspects of Residual Networks and Densely Connected CNNs (DenseNets, \cite{huang_densely_2017}). 
As described above, ResNets employ skip-connections within each single residual block to enable better information flow. 
DenseNets in turn introduce these paths between all residual blocks, concatenating input and output features from different network levels. 
The authors analyze both architectures and discover that residuals path implicitly favor features re-usage, but are limited in exploring new features, whereas DenseNets are better at exploring new features but suffer from heavy feature redundancy. 
This observation leads to the proposal of a combined architecture that is shown to outperform both ResNets and DenseNets in a variety of computer vision tasks.

\subsection{Decoders}\label{decoders}
It is common in semantic segmentation architectures to decouple the design of the encoder and the decoder, as this provides a better understanding of each component \cite{wojna_devil_2019}. 
In addition, it is usual to adopt pretrained encoders as described above, which leaves some design freedom to focus on specific characteristics of encoders that may allow for better reconstruction and upsampling techniques. 
We review below the three encoder architectures that we will be using in our experimental evaluation.

\subsubsection{U-Net decoder}
We first adopt the encoder branch of the widely popular U-Net architecture, one of the first CNNs designed for image segmentation, with particular success in biomedical imaging \cite{ronneberger_u-net_2015}. 
U-Net was among the first architecture designs to introduce the idea of an encoding (downsampling) branch followed by a decoding (upsampling) branch, with skip connections mapping intermediate activation from one branch to the other. 
This was shown to improve gradient flow during backpropagation and iterative weight updating, being also useful for recovering high-resolution information from the downsampling branch and adding it to the upsampling activations. 
The decoder branch of the U-Net features transposed convolutions as the most distinctive characteristic.

\subsubsection{DeepLab}

The first version of this architecture was presented in \cite{chen_deeplab_2018}, with the main feature being that atrous convolution that were leveraged to gain control on the resolution at which feature responses are computed by the network. 
Further improvements have been introduced to this architecture later on, and in this work we adopt DeepLab V3 \cite{chen_rethinking_2017}, which differs form DeepLab V1 on that atrous spatial pyramid pooling and batch-norm layers are adopted to increase long-range information modeling and improve training efficiency.

\subsubsection{Feature-Pyramid Networks}

FPNs were introduced in \cite{lin_feature_2017}. 
The authors proposed to build segmentation predictions not only in the last layer of the decoder but also at intermediate layers. 
These predictions range from low to high-resolution, and are combined into a single segmentation only at the end of the forward pass, in order to compare it to the ground-truth. 
This multi-scale approach is shown to effectively re-use information from all layers of the decoder, and results in increased performance.

\subsection{Training Details}

All the models trained in this work follow the same protocol. 
We optimize network weights to minimize the cross-entropy loss using standard Stochastic Gradient Descent with a learning rate of $l=0.01$ and a batch-size of $4$. 
The learning rate is decayed following a cosine law from its initial value to $l=1e-8$ during $25$ epochs, which defines a training cycle. 
We then repeat this process for $20$ cycles, restarting the learning back at the beginning of each cycle.
Images are re-scaled to $640\times512$, which respects the predominant rectangular aspect ratio in most polyp segmentation datasets, and during training they are augmented with standard techniques (random rotations, vertical/horizontal flipping, contrast/saturation/brightness changes). 
The mean Dice score is monitored on a separate validation set and the best performing model is kept for testing purposes. 
For testing, we generate four different versions of each image by horizontal/vertical flipping, predict on each of them, and average the results.

%% file: 3_results.tex
\begin{table}[t]  %
	\renewcommand{\arraystretch}{1.3}	
	\centering
\setlength\tabcolsep{1pt}	
\begin{tabular}{lc}
\textbf{Dataset}    &  \textbf{Characteristics \& Challenges}    \\
\midrule
KVasir-Seg \cite{jha_kvasir-seg_2020}         				      &   
\makecell{Large scale (n=1000), diverse, includes multi-polyp cases\\Varying resolution from $487\times332$ to $1920\times1072$ pixels} \\
\midrule
CVC-ClinicDB \cite{bernal_wm-dova_2015}       				             &   
\makecell{n=612 images of resolution $388\times284$ from 31 sequences\\Different polyp categories and sizes, substantial specularities} \\ %
\midrule
CVC-ColonDB \cite{bernal_towards_2012}      				                 &   
\makecell{n=300 images from 15 short sequences, $574\times500$ resolution\\Great variability in types of polyp and appearances} \\
\midrule
ETIS-LaribPolypDB \cite{silva_toward_2014}              				             &   
\makecell{n=196 still images with a resolution of $1225\times966$ \\ Contains 44 different polyps from 34 sequences} \\
\midrule
EndoScene \cite{gao_benchmark_2017}    			                 &   
\makecell{n=912, combination of CVC-ColonDB + CVC-ClinicDB\\Images of varying resolutions, only part of the test set used} \\
\bottomrule\\[0.025cm]
\end{tabular}
\caption{Description of each of the datasets considered in this paper. The first two are employed for training purposes. We use the five of them to assess performance on images similar/different to the training set.}
\label{tab_datasets}
\end{table}%

\section{Experimental Results}

This section describes our experimental analysis on the performance of the proposed approach to polyp segmentation. 
We first describe the data and the performance metrics used to validate our technique. 
We then develop an analysis of the accuracy of different combinations of encoders and decoders, comparing the performance on the Kvasir dataset~\cite{jha_kvasir-seg_2020} of a standard encoder-decoder network with that of its double encoder-decoder counterpart. 
Next we re-train the best model in a large-scale experiment and comprehensively analyze its performance when compared to other recent polyp segmentation methods on five different databases.

\subsection{Data and Evaluation Metrics}

It is certainly complicated to compare different approaches for polyp segmentation, as no clear evaluation protocol has emerged from the literature. This would involve clearly defining training and test subsets from the different publicly available datasets, as well as defining proper evaluation metrics.

In order to provide a fair comparison, we propose the following experimental setup. 
For the first set of experiments, we consider the KVasir-seg database~\cite{jha_kvasir-seg_2020}, which we split into 90\% training and 10\% test images. 
In this dataset, we carry out ablation experiments to find out which is the best encoder-decoder combination, and to demonstrate that double encoder-decoders offer superior performance. 
For the second part of this section, where we compare our approach against the state-of-the-art in a large-scale experiment, we follow closely the data source definitions provided in \cite{fan_pranet_2020}. This involves up to five different databases that are described in Table~\ref{tab_datasets}. 
The considered training set is the union of two subsets of KVasir-seg~\cite{jha_kvasir-seg_2020} and CVC-Clinic-DB~\cite{bernal_wm-dova_2015}, containing 90\% of the data. 
From this training set, a validation set is separated before training is carried out. 
To evaluate the performance of our models, we use the remaining 10\% of both datasets as test sets, which allow us to assess performance when test data is similar to training data. 
In addition, we use the three other datasets shown in Table~\ref{tab_datasets} to assess performance on data that comes from a different source: CVC-ColonDB~\cite{bernal_towards_2012}, ETIS-LaribDB~\cite{silva_toward_2014}, EndoScene~\cite{gao_benchmark_2017}. 
Note that the training/test splits are exactly the same\footnote{All data, training/test splits, and results of other compared methods were directly downloaded from \url{https://github.com/DengPingFan/PraNet}} as in \cite{fan_pranet_2020}. 
In particular, we only use the part of the EndoScene test set that does not belong to CVC-ClinicDB, which is used for training, referred to as CVC-300 below.
All other performance results are also extracted from \cite{fan_pranet_2020}, which ensures a correct comparison of all approaches.

Regarding performance metrics, a consensus has not been reached yet in the literature \cite{sanchez-peralta_eigenloss_2020}. We therefore follow the recommendation in \cite{jha_kvasir-seg_2020} and compute Dice and Intersection over Union (IoU) scores on each binary prediction of a given test set, and report their averages across all images. 
Note that this is different from \cite{fan_pranet_2020}, where the probabilistic (grayscale) predictions are used to derive an overall Dice and IoU score: if predictions are considered to be discretized in 8-bits, for each threshold in a predefined range $\{0,1,...,255\}$ a score is computed. Then the mean score is kept for each image, and the average of these is reported as the final performance. 
This is different from conventional performance metrics reported in other papers, and we prefer to follow the simpler approach of employing binary predictions to compute a single Dice and IoU score, as the binary segmentation is the most relevant outcome of a polyp segmentation pipeline, and it does not appear to be suitable to test a method on thresholds that are never going to be used in clinical practice. 
Nevertheless, as in \cite{fan_pranet_2020}, in order to also assess the quality of the probabilistic predictions, we report Mean Absolute Error computed as the average of the absolute value of pixelwise differences between the binary ground-truth and the grayscale predicted segmentation.

In order to generate binary segmentations from probabilistic predictions and calculate Dice/IOU scores, we use a simple adaptive thresholding algorithm \cite{otsu_threshold_1979}, which is a popular approach in polyp segmentation \cite{akbari_polyp_2018}, followed by morphological hole filling. 
The same binarization scheme is applied to the grayscale predictions of the compare methods, which were provided by the authors of \cite{fan_pranet_2020}.

\subsection{Autoencoders vs Double Autoencoders}

\begin{table*}[t]  %
	\renewcommand{\arraystretch}{1.4}	
	\centering
\setlength\tabcolsep{3pt}	
\begin{tabular}{l ccccccccc}
 \textbf{Decoder} $\boldsymbol{\rightarrow}$ & \multicolumn{3}{c}{\textbf{U-Net} \cite{ronneberger_u-net_2015}} & \multicolumn{3}{c}{\textbf{DeepLab} \cite{chen_rethinking_2017}} & \multicolumn{3}{c}{\textbf{FP-Net} \cite{lin_feature_2017}} \\
 \cmidrule(lr){1-1} \cmidrule(lr){2-4} \cmidrule(lr){5-7} \cmidrule(lr){8-10} 
\textbf{Encoder} $\boldsymbol{\downarrow}$        &  DICE &  IOU  &  MAE   &  DICE &  IOU  &    MAE   &  DICE  &   IOU  &  MAE     \\
\midrule
\textbf{MobileNet} \cite{sandler_mobilenetv2_2018}         & 88.59 & 82.87 &  3.65  &  88.87 & 82.88 &  3.79    &  89.15 &  83.14 &  3.61 \\
\textbf{MobileNet$\times2$}  & 88.80 & 83.24 &  3.68  &  90.23 & 84.80 &  3.59    &  90.73 &  85.53 &  3.17 \\
\hdashline[2pt/5pt]
\textbf{Perf. Diff.}         & \textbf{\Plus 0.21} & \textbf{\Plus 0.37} &  \Plus0.03 &  \textbf{\Plus 1.36} & \textbf{\Plus 1.96} &  \textbf{-0.20}   &  \textbf{\Plus 1.58} &  \textbf{\Plus2.39} & \textbf{-0.44} \\
\midrule 
\textbf{Resnet34} \cite{he_deep_2016}           & 89.72 & 84.28 &  3.24  & 89.26  & 83.66 &  3.31    &  89.90 &  84.76 &  3.00   \\ 
\textbf{Resnet34$\times2$}   & 90.13 & 84.77 &  3.09  & 90.39  & 85.30 &  3.12    &  90.70 &  85.55 &  3.01    \\
\hdashline[2pt/5pt]
\textbf{Perf. Diff.}         & \textbf{\Plus 0.41} & \textbf{\Plus 0.49} &  \textbf{-0.15} &  \textbf{\Plus 1.13} & \textbf{\Plus 1.64} &  \textbf{-0.19}  &  \textbf{\Plus0.80} &  \textbf{\Plus0.79} & \Plus0.01 \\
\midrule 
\textbf{DPN} \cite{chen_dual_2017}                & 89.72 & 83.91 &  3.19  & 90.23  & 84.83 &  3.25    &  89.96 &  84.61 &  3.04    \\
\textbf{DPN$\times2$}        & 90.21 & 85.10 &  3.25  & 91.21  & 86.16 &  2.84    &  91.97 &  87.05 &  2.65    \\
\hdashline[2pt/5pt]
\textbf{Perf. Diff.}         & \textbf{\Plus0.49} & \textbf{\Plus1.19} &  \Plus0.06 &  \textbf{\Plus0.98} & \textbf{\Plus1.33} &  \textbf{-0.39}  &  \textbf{\Plus2.01} &  \textbf{\Plus2.44} & \textbf{-0.39} \\
\bottomrule\\[0.05cm]
\end{tabular}
\caption{Performance analysis of different combinations of pretrained encoders and decoders on the Kvasir databaset. Bold numbers indicate performance improvements of double encoder-decoders upon their single encoder-decoder counterparts.}
\label{tab_results1}
\end{table*}%

We first train each of the combinations of pretrained encoders and decoders that we described in Sections \ref{encoders} and \ref{decoders} on the K-Vasir dataset. 
Our aim with this set of experiments is to investigate if using double encoder-decoders brings performance improvements.

Table \ref{tab_results1} reports the results of our experiments with different encoders and decoders. 
The main pattern arising is that double encoder-decoders do provide a performance boost in every considered case. 
Although the improvement is varying, there is no single case in which single encoder-decoder networks outperform their doubled counterparts. 
In addition, we can see that Feature-Pyramid networks (FP-Net) are the best decoder for this particular problem, surpassing the performance of U-Net and DeepLab for whichever considered encoder network. 
For the particular case of FP-Net encoders, it is also interesting to note that the lightweight MobileNet encoder achieves a better performance when compared to the heavier ResNet34 encoder. 
It is also worth noting that Dual Path Networks (DPN) stand out as the best pretrained encoder for all the considered decoders. 
In the remaining of this section, we select the DPN+FP-Net combination, and evaluate the performance of a double encoder-decoder network against other recent polyp segmentation techniques.

\subsection{Comparison with Recent Techniques}

In this section we re-train the best double encoder-decoder network (referred to as DPN68$\times2$) on the dataset employed in \cite{fan_pranet_2020} following the same train/test split. 
This training set derived from the Kvasir and CVC-Clinic databases is used to optimize the double DPN68$\times2$ architecture, and we conduct two types of experiments. 
First, we test the resulting model on data similar to the training distribution by computing performance (separately) on the KVasir and the CVC-Clinic test sets. 
Second, we use the ColonDB and ETIS databases, which contain relatively dissimilar colonoscopic frames, to assess the generalization ability of our approach. 
Following \cite{fan_pranet_2020}, we also include in this experiment the CVC-300 database, although this dataset contains part of the test set of CVC-Clinic, and it represents therefore an easier problem. 

In both cases, we compare the performance of our technique with a) two popular medical image segmentation architectures, namely U-Net \cite{ronneberger_u-net_2015} and its extensions U-Net++ \cite{zhou_unet_2020}, and b) three architectures specifically developed for  polyp segmentation techniques, ResUnet++ \cite{jha_resunet_2019}, SFA \cite{fang_selective_2019}, and PraNet \cite{fan_pranet_2020}. 

\begin{table*}[t]  %
	\renewcommand{\arraystretch}{1.3}	
	\centering
\setlength\tabcolsep{5pt}	
\begin{tabular}{l cccccc}
  & \multicolumn{3}{c}{\textbf{KVasir}} & \multicolumn{3}{c}{\textbf{CVC-Clinic}} \\
\cmidrule(lr){2-4} \cmidrule(lr){5-7}
				                &  DICE &  IOU  &  MAE   &  DICE &  IOU  &    MAE       \\
\midrule
\textbf{U-Net / MICCAI'15 \cite{ronneberger_u-net_2015}}     
								& 83.00 & 75.91 &  5.47  &  84.39 & 77.70 &  1.92      \\
\textbf{ResUNet-mod}\cite{jha_resunet_2019}           
						        & 79.09 & 42.87 &  ----- &  77.88 & 45.45 &  -----    \\
\textbf{ResUNet++}  \cite{jha_resunet_2019}
							    & 81.33 & 79.27 &  ----- &  79.55 & 79.62 &  -----    \\
\textbf{SFA / MICCAI'19 \cite{fang_selective_2019}}      
							    & 73.10 & 61.87 &  7.54  &  70.61 & 61.26 &  4.16      \\
\textbf{U-Net++ / TMI'20 \cite{zhou_unet_2020}} 
						        & 83.21 & 75.60 &  14.95 &  81.20 & 74.67 &  7.97     \\
\textbf{PraNet / MICCAI'20 \cite{fan_pranet_2020}} 
							    & 90.37 & 84.79 &  2.96  &  90.72 & 85.89 &  \textbf{0.93}    \\
\midrule 							    
\textbf{DPN68$\times2$}        & \textbf{91.70} & \textbf{86.74}  &  \textbf{2.68}  &  \textbf{91.61} & \textbf{86.65} &  1.42    \\
\bottomrule\\[0.05cm]
\end{tabular}
\caption{Performance of the best double encoder-decoder model (DPN68$\times2$) compared to other approaches on the test set of the KVasir-Seg and CVC-Clinic databases. Models were learned on the combination of the training set of both datasets.}
\label{tab_results2}
\end{table*}%

Results of the first of our experiments are reported in Table \ref{tab_results2}. 
It can be seen that the DPN68$\times2$ architecture improves the performance of all other approaches, achieving top performance by a wide margin with respect to most considered methods. The second best approach is the recently introduced PraNet CNN, which attains a performance relatively close to the one of DPN68$\times2$, although still slightly inferior.

It is however on our second experiment that reveals most remarkable performance differences between our approach and PraNet and the other considered techniques. 
The results displayed in Table \ref{tab_results3} show that the generalization capability of our DPN68$\times2$ model is much larger than previous approaches: when compared to the second best model (again PraNet), it attains an increase 5.29 in DICE score in the CVC-ColonDB and 12.07 Dice score in the ETIS databases. 
This margin is even larger for other methods, which clearly signals to DPN68$\times2$ being substantially more competitive when using unseen data that is far away from the training distribution. 
On the CVC-300 database, which is more similar to the training data, DPN68$\times2$ still surpasses all other methods, although with a smaller difference.

\begin{table*}[h]  %
	\renewcommand{\arraystretch}{1.3}	
	\centering
\setlength\tabcolsep{1.5pt}	
\begin{tabular}{l ccccccccc}
  & \multicolumn{3}{c}{\textbf{ColonDB}} & \multicolumn{3}{c}{\textbf{ETIS}} & \multicolumn{3}{c}{\textbf{CVC-300}} \\
\cmidrule(lr){2-4} \cmidrule(lr){5-7} \cmidrule(lr){8-10}
				                       &  DICE &  IOU  &  MAE   &  DICE  &  IOU  &    MAE    &  DICE  &  IOU  &    MAE    \\
\midrule
\textbf{U-Net / MICCAI'15} 
                                       & 53.91 & 46.28 &  5.86  &  44.02 & 36.72 &  3.62     &  73.99 & 65.30 &  2.21  \\
\textbf{SFA / MICCAI'19 }
						               & 45.90 & 33.83 &  7.46  &  29.88 & 21.82 &  4.55     &  46.90 & 32.95 &  6.51  \\
\textbf{U-Net++ / TMI'20 }              
& 51.73 & 43.37 &  7.42  &  48.05 & 40.15 &  4.52     &  74.79 & 65.89 &  3.36 \\
\textbf{PraNet / MICCAI'20 }   		   
								       & 72.42 & 65.11 &  4.31  &  64.12 & 57.98 &  3.11     &  87.63 & 80.35 &  0.99  \\
\midrule 	
\textbf{DPN68$\times2$}                & \textbf{77.71} & \textbf{69.81} &  \textbf{3.97}  &  \textbf{76.19} & \textbf{68.43} &  \textbf{1.68} &  \textbf{89.15} & \textbf{82.56} &  \textbf{0.90} \\
\bottomrule\\[0.05cm]
\end{tabular}
\caption{Performance of the best double encoder-decoder model (DPN68$\times2$) compared to other approaches on the ColonDB, ETIS, and CVC-300 databases.}
\label{tab_results3}
\end{table*}%

\begin{figure*}[h]
\centering
\subfloat[]{\includegraphics[width = 0.20\textwidth]{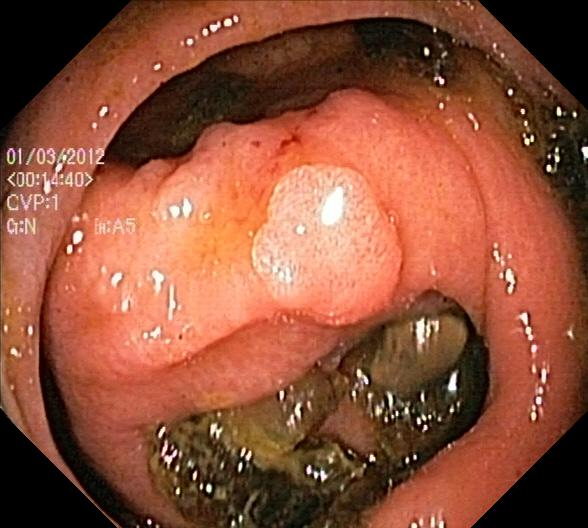}
\label{fig_seg_1}}
\hfil
\subfloat[]{\includegraphics[width = 0.20\textwidth]{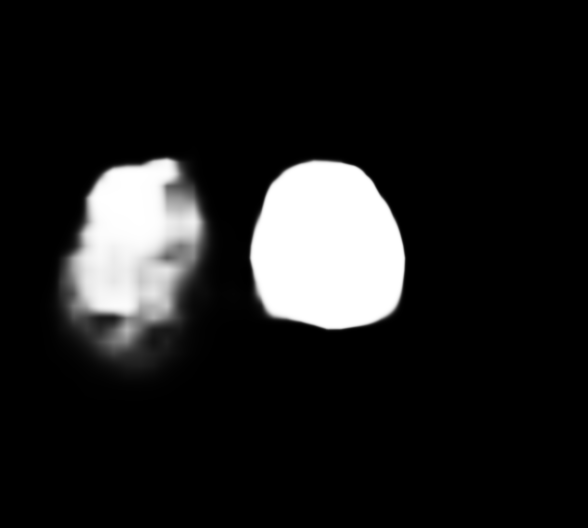}
\label{fig_seg_2}}
\hfil
\subfloat[]{\includegraphics[width = 0.20\textwidth]{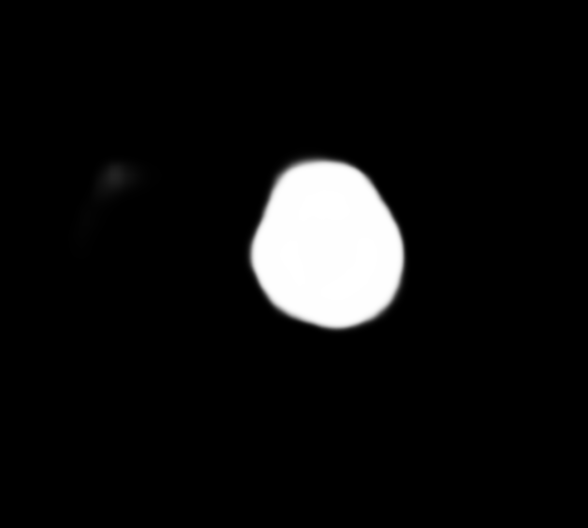}
\label{fig_seg_3}}
\hfil
\subfloat[]{\includegraphics[width = 0.20\textwidth]{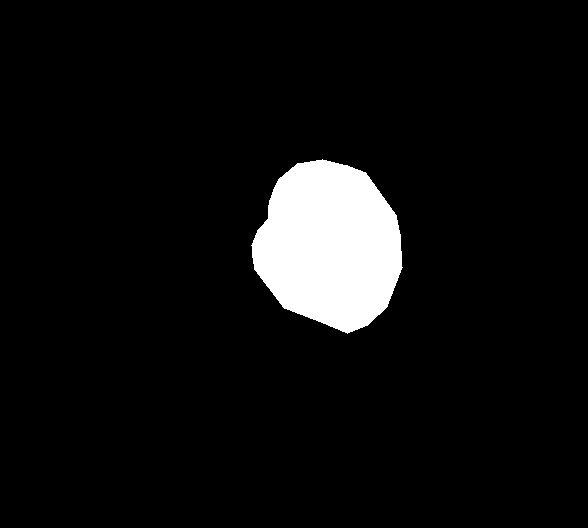}
\label{fig_seg_4}}

\subfloat[]{\includegraphics[width = 0.20\textwidth]{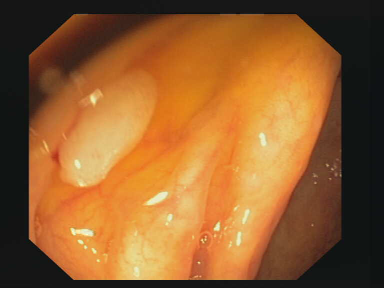}
\label{fig_seg_5}}
\hfil
\subfloat[]{\includegraphics[width = 0.20\textwidth]{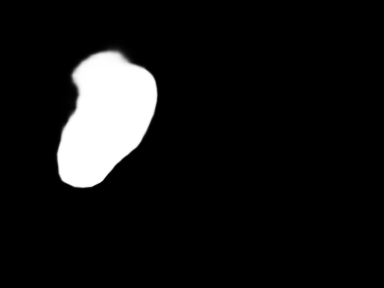}
\label{fig_seg_6}}
\hfil
\subfloat[]{\includegraphics[width = 0.20\textwidth]{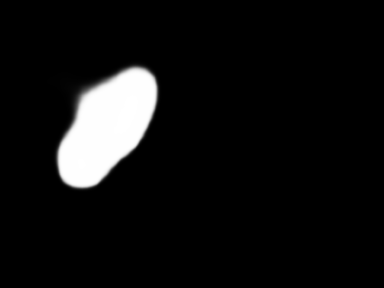}
\label{fig_seg_7}}
\hfil
\subfloat[]{\includegraphics[width = 0.20\textwidth]{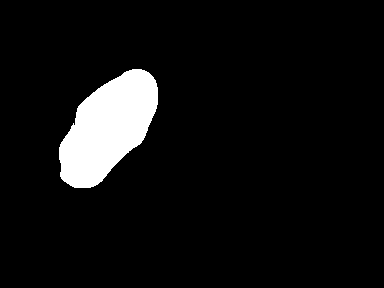}
\label{fig_seg_8}}
\caption{(a), (e): Polyps extracted from the Kvasir-Seg and CVC-Clinic databases. Segmentations produced by (b), (f): PraNet, (c), (g): DPN68$\times2$. (d), (h): Ground-truth}
\label{fig_polyps_seg}
\end{figure*}

\subsection{Qualitative Analysis}
In this section, we visually analyze some of the segmentation results obtained on the different test data when compared with other approaches.

\begin{figure*}[t]
\centering
\subfloat[]{\includegraphics[width = 0.22\textwidth]{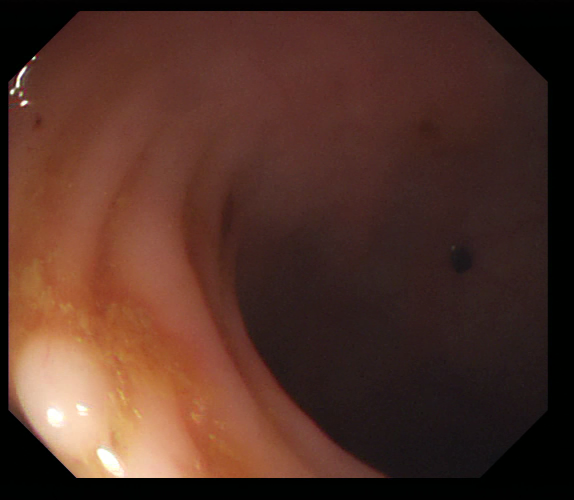}
\label{fig_seg_a}}
\hfil
\subfloat[]{\includegraphics[width = 0.22\textwidth]{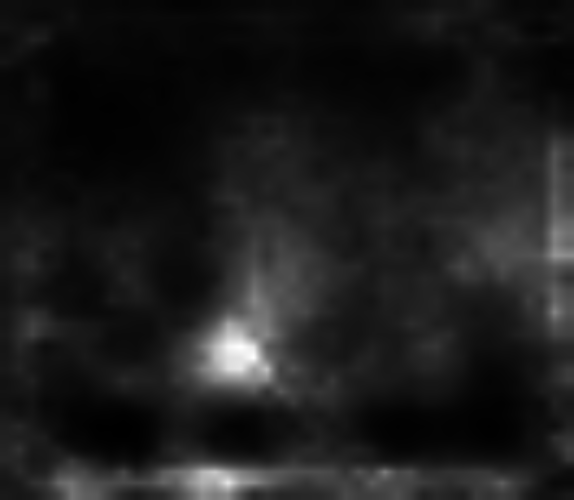}
\label{fig_seg_b}}
\hfil
\subfloat[]{\includegraphics[width = 0.22\textwidth]{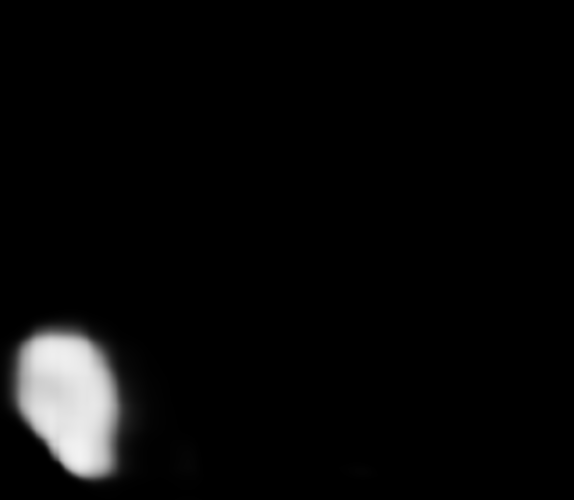}
\label{fig_seg_c}}
\hfil
\subfloat[]{\includegraphics[width = 0.22\textwidth]{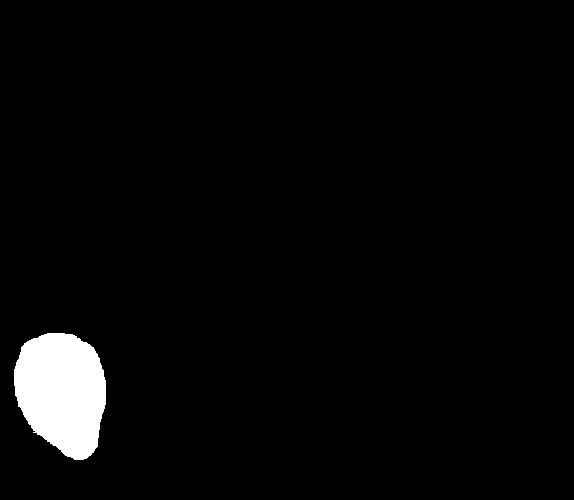}
\label{fig_seg_d}}

\subfloat[]{\includegraphics[width = 0.22\textwidth]{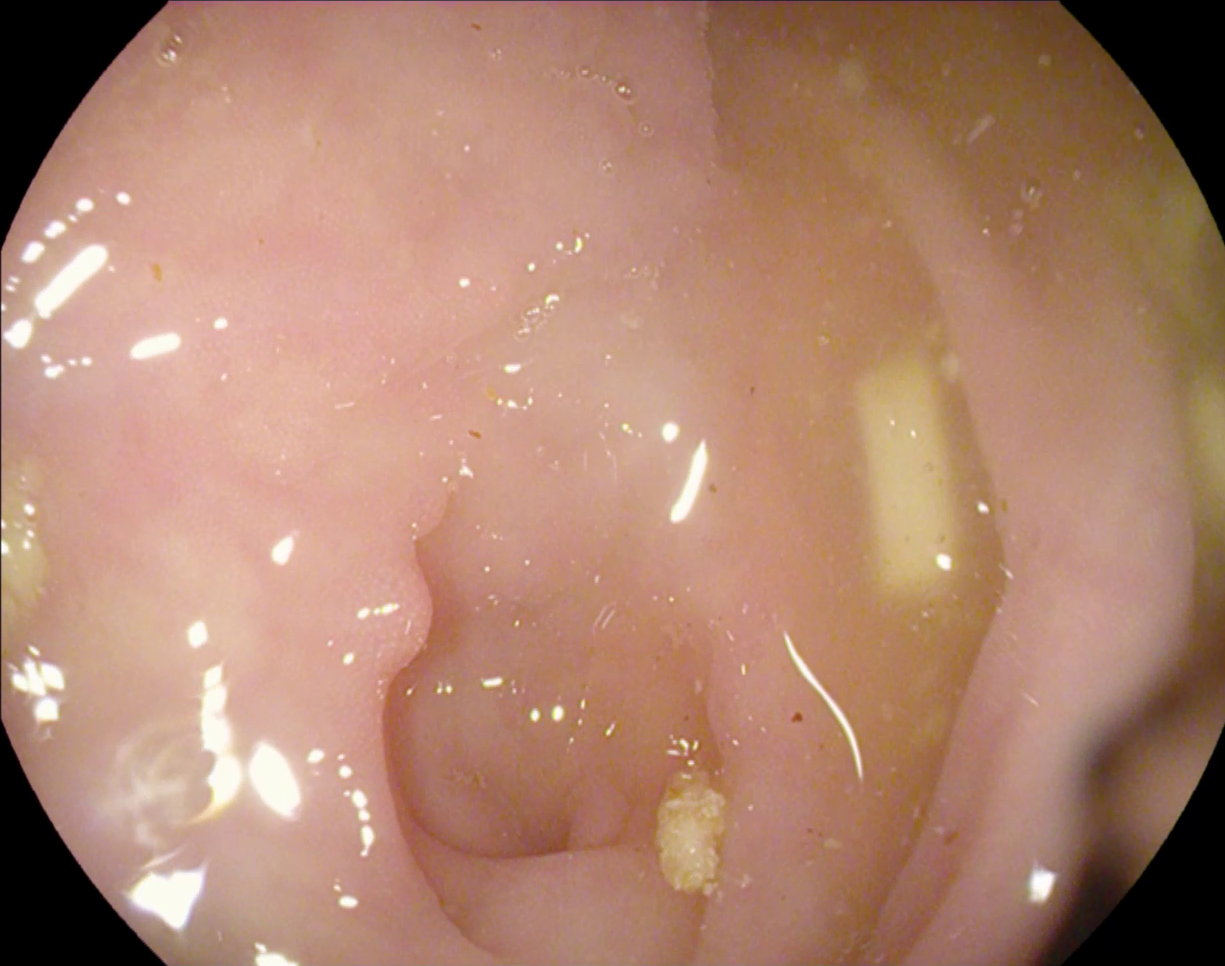}
\label{fig_seg_e}}
\hfil
\subfloat[]{\includegraphics[width = 0.22\textwidth]{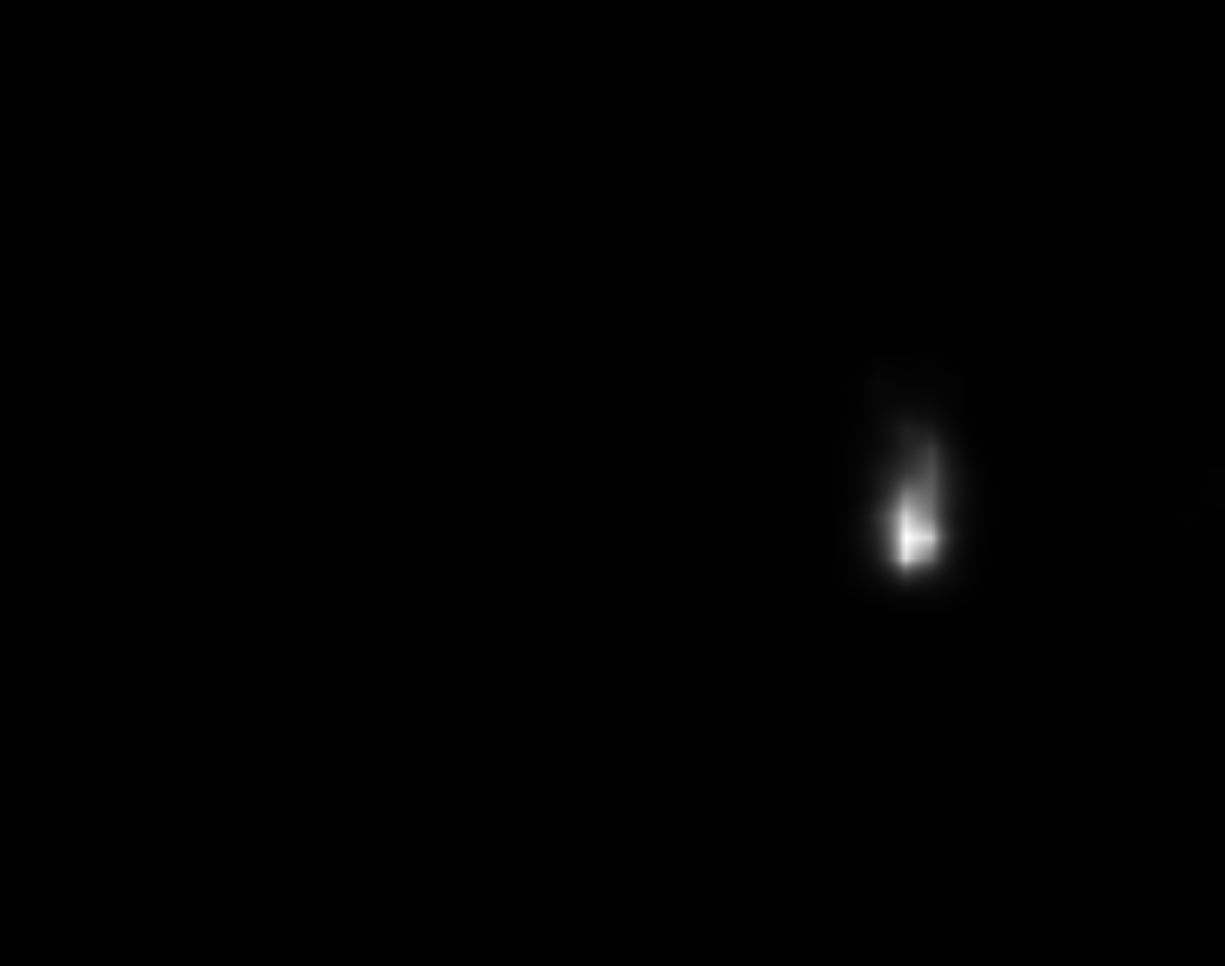}
\label{fig_seg_f}}
\hfil
\subfloat[]{\includegraphics[width = 0.22\textwidth]{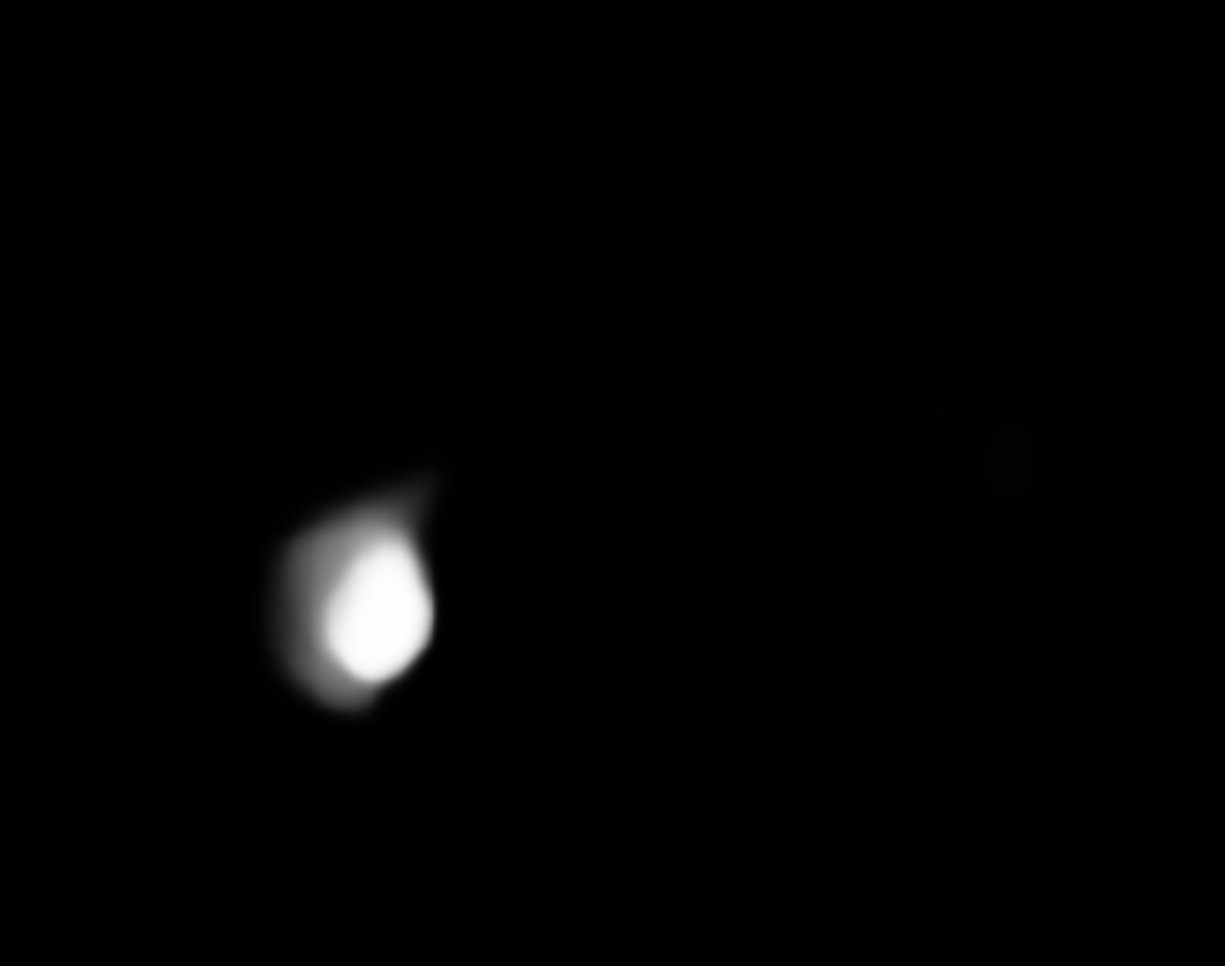}
\label{fig_seg_g}}
\hfil
\subfloat[]{\includegraphics[width = 0.22\textwidth]{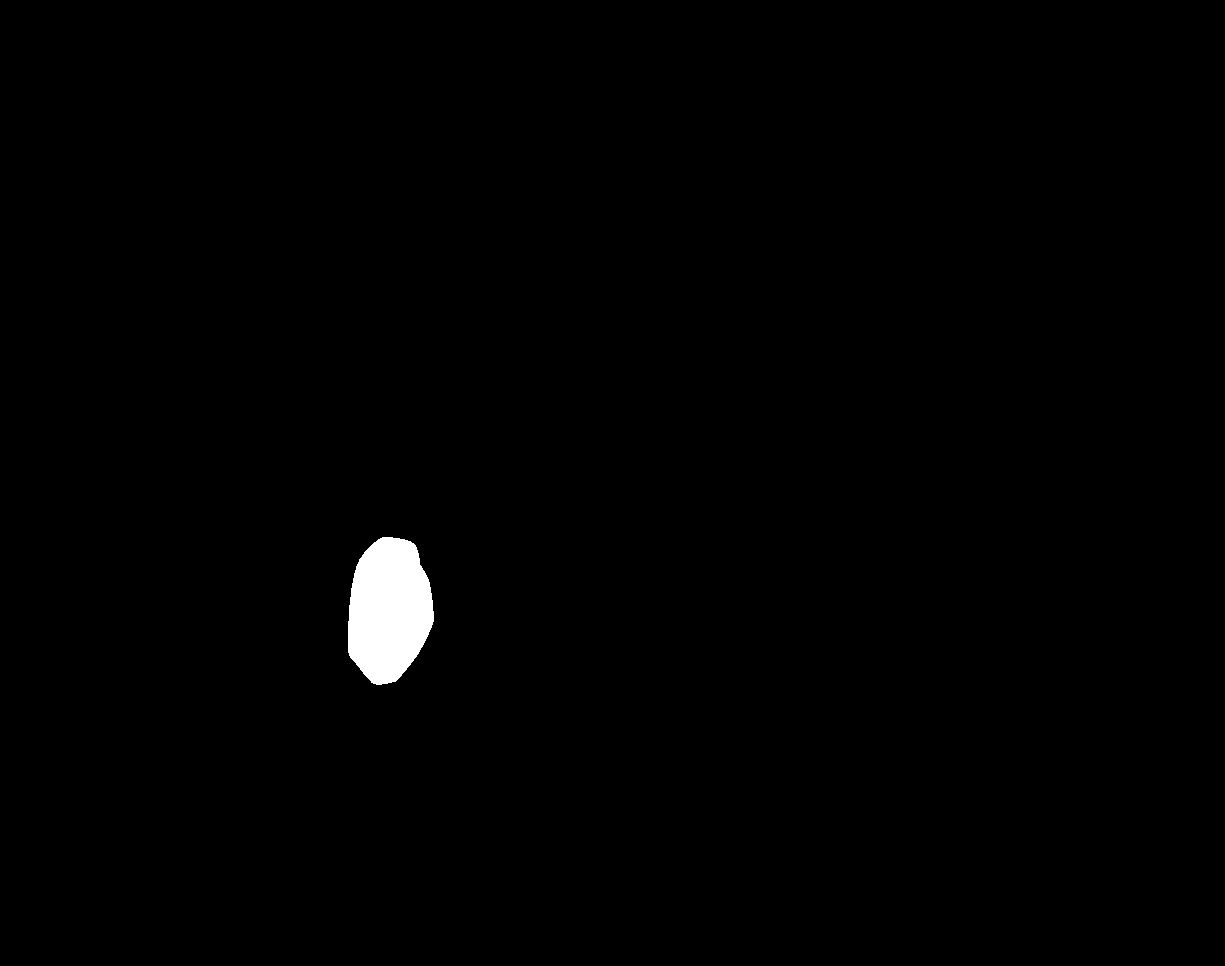}
\label{fig_seg_h}}
\caption{(a), (e): Polyps extracted from the CVC-ColonDB and ETIS databases. Segmentations produced by (b), (f): PraNet, (c), (g): DPN68$\times2$. (d), (h): Ground-truth}
\label{fig_polyps_seg2}
\end{figure*}

Fig. \ref{fig_polyps_seg} shows two challenging cases of polyps extracted from the Kvasir and CVC-Clinic databases (test set associated to the training set), while Fig. \ref{fig_polyps_seg2} displays two polyps extracted from the test set of ColonDB and ETIS (dissimilar data). 
We also show the results produced by the second best technique, PraNet. 

In the Kvasir example, it can be appreciated that DPN68$\times2$ succeeds to properly segment the polyp appearing in the center of the frame in Fig. (\ref{fig_seg_1}), whereas PraNet is probably confused by the specularities present in the image and oversegments this example, as shown in Figs. (\ref{fig_seg_2}) and (\ref{fig_seg_3}). 
As for the CVC-Clinic example in Fig. (\ref{fig_seg_5}), the polyp appearing to the left of the frame is reasonably segmented by both methods. 
However, the confounding artifact above the polyp due to optical blurring is wrongly segmented by PraNet as part of the anomaly, as seen in (Fig. \ref{fig_seg_6}); DPN68$\times2$ is also slightly confused, but it assigns much less probability to the pixels pixels related to the blurring artifact in Fig. (\ref{fig_seg_7}).

A challenging example extracted from the CVC-ColonDB is shown in the top row of Fig. \ref{fig_polyps_seg2}. 
This low-contrast frame leads to a wrong prediction in the PraNet case shown in Fig. (\ref{fig_seg_b}), whereas the DPN68$\times2$ model has no trouble in accurately delineating the borders of the polyp in Fig. (\ref{fig_seg_c}).
Finally, an example from the ETIS database is shown in the bottom row of Fig. \ref{fig_polyps_seg2}. 
This is a hardly visible polyp, and PraNet erroneously locates it to the right of the image, while DPN68$\times2$, even if producing some oversegmenting, avoids this confusion.

%% file: 4_discussion.tex
\section{Discussion and Conclusion}
In this work we have analyzed double decoder-encoder networks for the task of polyp segmentation on endoscopic images. 
The encoder of the first network generates an attention map that indicates the second network the most interesting areas of the image. 
Double encoder-decoders leverage state-of-the-art encoder pretrained large natural image databases. 
We have shown that they provide a clear advantage over their single encoder-decoder counterparts. 
Comprehensive experimental results also show that a double encoder-decoder with a DPN encoder and a FP-Net decoder attains state-of-the-art on common benchmarks, with particularly good results on data that is visually dissimilar to the training images. 
Future steps involve the investigation of more efficient combination of the two building blocks of double encoder-decoders. 
It is reasonable to expect that smaller networks can already generate a suitable attention map that can be employed by the second network to achieve similar results at a reduced computational load.

%% file: ms.bbl
\begin{thebibliography}{10}
\providecommand{\url}[1]{\texttt{#1}}
\providecommand{\urlprefix}{URL }
\providecommand{\doi}[1]{https://doi.org/#1}

\bibitem{ahn_miss_2012}
Ahn, S.B., Han, D.S., Bae, J.H., Byun, T.J., Kim, J.P., Eun, C.S.: The {Miss}
  {Rate} for {Colorectal} {Adenoma} {Determined} by {Quality}-{Adjusted},
  {Back}-to-{Back} {Colonoscopies}. Gut and Liver  \textbf{6}(1),  64--70 (Jan
  2012)

\bibitem{akbari_polyp_2018}
Akbari, M., Mohrekesh, M., Nasr-Esfahani, E., Soroushmehr, S.R., Karimi, N.,
  Samavi, S., Najarian, K.: Polyp {Segmentation} in {Colonoscopy} {Images}
  {Using} {Fully} {Convolutional} {Network}. In: 2018 40th {Annual}
  {International} {Conference} of the {IEEE} {Engineering} in {Medicine} and
  {Biology} {Society} ({EMBC}). pp. 69--72 (Jul 2018), iSSN: 1558-4615

\bibitem{badrinarayanan_segnet_2017}
Badrinarayanan, V., Kendall, A., Cipolla, R.: {SegNet}: {A} {Deep}
  {Convolutional} {Encoder}-{Decoder} {Architecture} for {Image}
  {Segmentation}. IEEE Transactions on Pattern Analysis and Machine
  Intelligence  \textbf{39}(12),  2481--2495 (Dec 2017)

\bibitem{bernal_towards_2012}
Bernal, J., Sánchez, J., Vilariño, F.: Towards automatic polyp detection with
  a polyp appearance model. Pattern Recognition  \textbf{45}(9),  3166--3182
  (Sep 2012)

\bibitem{bernal_wm-dova_2015}
Bernal, J., Sánchez, F.J., Fernández-Esparrach, G., Gil, D., Rodríguez, C.,
  Vilariño, F.: {WM}-{DOVA} maps for accurate polyp highlighting in
  colonoscopy: {Validation} vs. saliency maps from physicians. Computerized
  Medical Imaging and Graphics  \textbf{43},  99--111 (Jul 2015)

\bibitem{bernal_comparative_2017}
Bernal, J., Tajkbaksh, N., Sánchez, F.J., Matuszewski, B.J., Chen, H., Yu, L.,
  Angermann, Q., Romain, O., Rustad, B., Balasingham, I., Pogorelov, K., Choi,
  S., Debard, Q., Maier-Hein, L., Speidel, S., Stoyanov, D., Brandao, P.,
  Córdova, H., Sánchez-Montes, C., Gurudu, S.R., Fernández-Esparrach, G.,
  Dray, X., Liang, J., Histace, A.: Comparative {Validation} of {Polyp}
  {Detection} {Methods} in {Video} {Colonoscopy}: {Results} {From} the {MICCAI}
  2015 {Endoscopic} {Vision} {Challenge}. IEEE Transactions on Medical Imaging
  \textbf{36}(6),  1231--1249 (Jun 2017)

\bibitem{carneiro_deep_2020}
Carneiro, G., Zorron Cheng Tao~Pu, L., Singh, R., Burt, A.: Deep learning
  uncertainty and confidence calibration for the five-class polyp
  classification from colonoscopy. Medical Image Analysis  \textbf{62},  101653
  (May 2020)

\bibitem{chen_deeplab_2018}
Chen, L.C., Papandreou, G., Kokkinos, I., Murphy, K., Yuille, A.L.: {DeepLab}:
  {Semantic} {Image} {Segmentation} with {Deep} {Convolutional} {Nets},
  {Atrous} {Convolution}, and {Fully} {Connected} {CRFs}. IEEE Transactions on
  Pattern Analysis and Machine Intelligence  \textbf{40}(4),  834--848 (Apr
  2018)

\bibitem{chen_rethinking_2017}
Chen, L.C., Papandreou, G., Schroff, F., Adam, H.: Rethinking {Atrous}
  {Convolution} for {Semantic} {Image} {Segmentation}. arXiv:1706.05587 [cs]
  (Dec 2017), arXiv: 1706.05587

\bibitem{chen_dual_2017}
Chen, Y., Li, J., Xiao, H., Jin, X., Yan, S., Feng, J.: Dual path networks. In:
  Proceedings of the 31st {International} {Conference} on {Neural}
  {Information} {Processing} {Systems}. pp. 4470--4478. {NIPS}'17, Curran
  Associates Inc., Red Hook, NY, USA (Dec 2017)

\bibitem{fan_pranet_2020}
Fan, D.P., Ji, G.P., Zhou, T., Chen, G., Fu, H., Shen, J., Shao, L.: {PraNet}:
  {Parallel} {Reverse} {Attention} {Network} for {Polyp} {Segmentation}. In:
  Medical {Image} {Computing} and {Computer} {Assisted} {Intervention} –
  {MICCAI} 2020 (2020)

\bibitem{fang_selective_2019}
Fang, Y., Chen, C., Yuan, Y., Tong, K.y.: Selective {Feature} {Aggregation}
  {Network} with {Area}-{Boundary} {Constraints} for {Polyp} {Segmentation}.
  In: Medical {Image} {Computing} and {Computer} {Assisted} {Intervention} -
  {MICCAI} 2019. pp. 302--310. Lecture {Notes} in {Computer} {Science},
  Springer International Publishing, Cham (2019)

\bibitem{galdran_little_2020}
Galdran, A., Anjos, A., Dolz, J., Chakor, H., Lombaert, H., Ayed, I.B.: The
  {Little} {W}-{Net} {That} {Could}: {State}-of-the-{Art} {Retinal} {Vessel}
  {Segmentation} with {Minimalistic} {Models}. arXiv:2009.01907  (Sep 2020)

\bibitem{guo_polyp_2020}
Guo, Y., Bernal, J., J.~Matuszewski, B.: Polyp {Segmentation} with {Fully}
  {Convolutional} {Deep} {Neural} {Networks}—{Extended} {Evaluation} {Study}.
  Journal of Imaging  \textbf{6}(7), ~69 (Jul 2020)

\bibitem{haggar_colorectal_2009}
Haggar, F.A., Boushey, R.P.: Colorectal {Cancer} {Epidemiology}: {Incidence},
  {Mortality}, {Survival}, and {Risk} {Factors}. Clinics in Colon and Rectal
  Surgery  \textbf{22}(4),  191--197 (Nov 2009)

\bibitem{he_deep_2016}
He, K., Zhang, X., Ren, S., Sun, J.: Deep {Residual} {Learning} for {Image}
  {Recognition}. In: 2016 {IEEE} {Conference} on {Computer} {Vision} and
  {Pattern} {Recognition} ({CVPR}). pp. 770--778 (Jun 2016), iSSN: 1063-6919

\bibitem{huang_densely_2017}
Huang, G., Liu, Z., Van Der~Maaten, L., Weinberger, K.Q.: Densely {Connected}
  {Convolutional} {Networks}. In: 2017 {IEEE} {Conference} on {Computer}
  {Vision} and {Pattern} {Recognition} ({CVPR}). pp. 2261--2269 (Jul 2017),
  iSSN: 1063-6919

\bibitem{jha_doubleu-net_2020}
Jha, D., Riegler, M., Johansen, D., Halvorsen, P., Johansen, H.D.:
  {DoubleU}-{Net}: {A} {Deep} {Convolutional} {Neural} {Network} for {Medical}
  {Image} {Segmentation}. 2020 IEEE 33rd International Symposium on
  Computer-Based Medical Systems (CBMS)  (2020)

\bibitem{jha_kvasir-seg_2020}
Jha, D., Smedsrud, P.H., Riegler, M.A., Halvorsen, P., de~Lange, T., Johansen,
  D., Johansen, H.D.: Kvasir-seg: {A} segmented polyp dataset. In:
  International {Conference} on {Multimedia} {Modeling}. pp. 451--462. Springer
  (2020)

\bibitem{jha_resunet_2019}
Jha, D., Smedsrud, P.H., Riegler, M.A., Johansen, D., Lange, T.D., Halvorsen,
  P., D.~Johansen, H.: {ResUNet}++: {An} {Advanced} {Architecture} for
  {Medical} {Image} {Segmentation}. In: 2019 {IEEE} {International} {Symposium}
  on {Multimedia} ({ISM}). pp. 225--2255 (Dec 2019)

\bibitem{jia_automatic_2020}
Jia, X., Mai, X., Cui, Y., Yuan, Y., Xing, X., Seo, H., Xing, L., Meng, M.Q.H.:
  Automatic {Polyp} {Recognition} in {Colonoscopy} {Images} {Using} {Deep}
  {Learning} and {Two}-{Stage} {Pyramidal} {Feature} {Prediction}. IEEE
  Transactions on Automation Science and Engineering  \textbf{17}(3),
  1570--1584 (Jul 2020)

\bibitem{kang_ensemble_2019}
Kang, J., Gwak, J.: Ensemble of {Instance} {Segmentation} {Models} for {Polyp}
  {Segmentation} in {Colonoscopy} {Images}. IEEE Access  \textbf{7},
  26440--26447 (2019)

\bibitem{li_iternet_2020}
Li, L., Verma, M., Nakashima, Y., Nagahara, H., Kawasaki, R.: {IterNet}:
  {Retinal} {Image} {Segmentation} {Utilizing} {Structural} {Redundancy} in
  {Vessel} {Networks}. In: The {IEEE} {Winter} {Conference} on {Applications}
  of {Computer} {Vision} ({WACV}) (Mar 2020)

\bibitem{lin_feature_2017}
Lin, T.Y., Dollár, P., Girshick, R., He, K., Hariharan, B., Belongie, S.:
  Feature {Pyramid} {Networks} for {Object} {Detection}. In: 2017 {IEEE}
  {Conference} on {Computer} {Vision} and {Pattern} {Recognition} ({CVPR}). pp.
  936--944 (Jul 2017), iSSN: 1063-6919

\bibitem{lui_new_2020}
Lui, T.K., Hui, C.K., Tsui, V.W., Cheung, K.S., Ko, M.K., aCC Foo, D., Mak,
  L.Y., Yeung, C.K., Lui, T.H., Wong, S.Y., Leung, W.K.: New insights on missed
  colonic lesions during colonoscopy through artificial intelligence–assisted
  real-time detection (with video). Gastrointestinal Endoscopy  (May 2020)

\bibitem{otsu_threshold_1979}
Otsu, N.: A {Threshold} {Selection} {Method} from {Gray}-{Level} {Histograms}.
  IEEE Transactions on Systems, Man, and Cybernetics  \textbf{9}(1),  62--66
  (Jan 1979)

\bibitem{ronneberger_u-net_2015}
Ronneberger, O., Fischer, P., Brox, T.: U-{Net}: {Convolutional} {Networks} for
  {Biomedical} {Image} {Segmentation}. In: Medical {Image} {Computing} and
  {Computer}-{Assisted} {Intervention} – {MICCAI} 2015. pp. 234--241. Lecture
  {Notes} in {Computer} {Science}, Springer International Publishing, Cham
  (2015)

\bibitem{sandler_mobilenetv2_2018}
Sandler, M., Howard, A., Zhu, M., Zhmoginov, A., Chen, L.C.: {MobileNetV2}:
  {Inverted} {Residuals} and {Linear} {Bottlenecks}. In: 2018 {IEEE}/{CVF}
  {Conference} on {Computer} {Vision} and {Pattern} {Recognition}. pp.
  4510--4520 (Jun 2018)

\bibitem{silva_toward_2014}
Silva, J., Histace, A., Romain, O., Dray, X., Granado, B.: Toward embedded
  detection of polyps in {WCE} images for early diagnosis of colorectal cancer.
  International Journal of Computer Assisted Radiology and Surgery
  \textbf{9}(2),  283--293 (Mar 2014)

\bibitem{simonyan_very_2015}
Simonyan, K., Zisserman, A.: Very {Deep} {Convolutional} {Networks} for
  {Large}-{Scale} {Image} {Recognition}. In: International {Conference} on
  {Learning} {Representations} (2015)

\bibitem{sanchez-peralta_deep_2020}
Sánchez-Peralta, L.F., Bote-Curiel, L., Picón, A., Sánchez-Margallo, F.M.,
  Pagador, J.B.: Deep learning to find colorectal polyps in colonoscopy: {A}
  systematic literature review. Artificial Intelligence in Medicine
  \textbf{108},  101923 (Aug 2020)

\bibitem{sanchez-peralta_eigenloss_2020}
Sánchez-Peralta, L.F., Picón, A., Antequera-Barroso, J.A., Ortega-Morán,
  J.F., Sánchez-Margallo, F.M., Pagador, J.B.: Eigenloss: {Combined}
  {PCA}-{Based} {Loss} {Function} for {Polyp} {Segmentation}. Mathematics
  \textbf{8}(8), ~1316 (Aug 2020)

\bibitem{tian_decoders_2019}
Tian, Z., He, T., Shen, C., Yan, Y.: Decoders {Matter} for {Semantic}
  {Segmentation}: {Data}-{Dependent} {Decoding} {Enables} {Flexible} {Feature}
  {Aggregation}. pp. 3126--3135 (2019)

\bibitem{gao_benchmark_2017}
Vázquez, D., Bernal, J., Sánchez, F.J., Fernández-Esparrach, G., López,
  A.M., Romero, A., Drozdzal, M., Courville, A.: A {Benchmark} for
  {Endoluminal} {Scene} {Segmentation} of {Colonoscopy} {Images}. Journal of
  Healthcare Engineering  \textbf{2017},  4037190 (Jul 2017)

\bibitem{wickstrom_uncertainty_2020}
Wickstrøm, K., Kampffmeyer, M., Jenssen, R.: Uncertainty and interpretability
  in convolutional neural networks for semantic segmentation of colorectal
  polyps. Medical Image Analysis  \textbf{60},  101619 (Feb 2020)

\bibitem{wojna_devil_2019}
Wojna, Z., Ferrari, V., Guadarrama, S., Silberman, N., Chen, L.C., Fathi, A.,
  Uijlings, J.: The {Devil} is in the {Decoder}: {Classification}, {Regression}
  and {GANs}. International Journal of Computer Vision  \textbf{127}(11),
  1694--1706 (Dec 2019)

\bibitem{xie_aggregated_2017}
Xie, S., Girshick, R., Dollár, P., Tu, Z., He, K.: Aggregated {Residual}
  {Transformations} for {Deep} {Neural} {Networks}. In: 2017 {IEEE}
  {Conference} on {Computer} {Vision} and {Pattern} {Recognition} ({CVPR}). pp.
  5987--5995 (Jul 2017), iSSN: 1063-6919

\bibitem{zhang_polyp_2018}
Zhang, R., Zheng, Y., Poon, C.C.Y., Shen, D., Lau, J.Y.W.: Polyp detection
  during colonoscopy using a regression-based convolutional neural network with
  a tracker. Pattern Recognition  \textbf{83},  209--219 (Nov 2018)

\bibitem{zhou_unet_2020}
Zhou, Z., Siddiquee, M.M.R., Tajbakhsh, N., Liang, J.: {UNet}++: {Redesigning}
  {Skip} {Connections} to {Exploit} {Multiscale} {Features} in {Image}
  {Segmentation}. IEEE Transactions on Medical Imaging  \textbf{39}(6),
  1856--1867 (Jun 2020)

\end{thebibliography}
